\documentclass[sigconf]{acmart}

\AtBeginDocument{%
  \providecommand\BibTeX{{%
    \normalfont B\kern-0.5em{\scshape i\kern-0.25em b}\kern-0.8em\TeX}}}

\copyrightyear{2024}
\acmYear{2024}
\setcopyright{acmlicensed}\acmConference[L@S '24]{Proceedings of the Eleventh ACM Conference on Learning @ Scale}{July 18--20, 2024}{Atlanta, GA, USA}
\acmBooktitle{Proceedings of the Eleventh ACM Conference on Learning @ Scale (L@S '24), July 18--20, 2024, Atlanta, GA, USA}
\acmDOI{10.1145/3657604.3662031}
\acmISBN{979-8-4007-0633-2/24/07}
\usepackage{multirow}
\author{Xinyi Lu}
\affiliation{
 \institution{University of Michigan}
 \department{Computer Science and Engineering}
 \city{Ann Arbor}
  \country{United States}
}
\email{lwlxy@umich.edu}

\author{Xu Wang}
\affiliation{
 \institution{University of Michigan}
\department{Computer Science and Engineering}
 \city{Ann Arbor}
 \country{United States}
}
\email{xwanghci@umich.edu}

\usepackage[normalem]{ulem}
\useunder{\uline}{\ul}{}

\begin{document}

\title{Generative Students: Using LLM-Simulated Student Profiles to Support Question Item Evaluation}



\begin{abstract}
Evaluating the quality of automatically generated question items has been a long standing challenge. 
In this paper, we leverage LLMs to simulate student profiles and generate responses to multiple-choice questions (MCQs). The generative students' responses to MCQs can further support question item evaluation.
We propose Generative Students, a prompt architecture designed based on the KLI framework. A generative student profile is a function of the list of knowledge components the student has mastered, has confusion about or has no evidence of knowledge of. We instantiate the Generative Students concept on the subject domain of heuristic evaluation. We created 45 generative students using GPT-4 and had them respond to 20 MCQs. We found that the generative students produced logical and believable responses that were aligned with their profiles. We then compared the generative students' responses to real students' responses on the same set of MCQs and found a high correlation. Moreover, there was considerable overlap in the difficult questions identified by generative students and real students. 
A subsequent case study demonstrated that an instructor could improve question quality based on the signals provided by Generative Students. 
\end{abstract}




\begin{teaserfigure}
    \centering
    \includegraphics[width=0.9\textwidth]{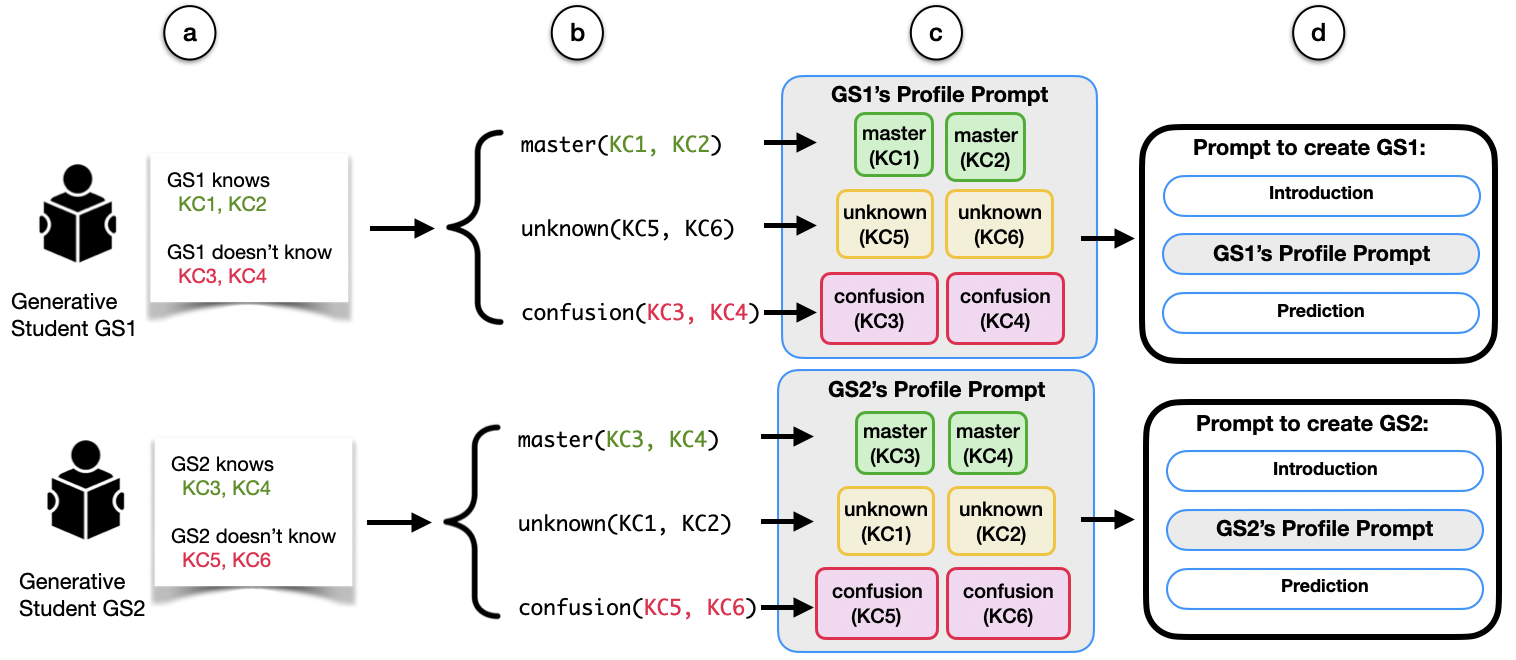}
    \caption{The design of the prompt architecture of Generative Students is based on the KLI framework, which uses knowledge components (KCs) to define the elements students are expected to learn. With the KCs identified for a given task (a), the generative student's profile is a function of the list of KCs the student has mastered, has confusion about, or has no evidence of knowledge of (b). Users can define master prompt, confusion prompt, and unknown prompt for a given task (c). This architecture thus supports automatic creation of diverse student profiles (d).}
    \label{fig:main}
\end{teaserfigure}

\maketitle
\section{Introduction}

Decades of educational research has shown the benefit of active learning \cite{chi2014icap, crouch2001peer, deslauriers2019measuring, koedinger2015learningbydoing}, one-on-one tutoring \cite{koedinger1997intelligent, bloom19842}, and deliberate practice \cite{ericsson1993deliberatepractice, ericsson2006influence} in improving students' learning outcomes. These theories highlight the benefit of providing students with hands-on problem solving and question answering opportunities to facilitate learning. It has been a long standing interest in the Learning at Scale and AI in Education communities to study effective question generation techniques \cite{alsubait2016ontology, kurdi2020systematic} to support the creation of high-quality question items at scale to enhance active learning, tutoring, and deliberate practice. Multiple-choice question creation is of particular interest because of their practical value in terms of ease of grading and automatic provision of feedback \cite{alsubait2016ontology, kurdi2020systematic, wang2022towards, wang2021seeing, wang2019upgrade, singh2021s}. 

Prior work has explored a variety of ways to support multiple-choice question creation for educational purposes, including crowdsourcing questions from students \cite{singh2021s, 10.1145/3313831.3376882}, perusing prior students' solutions and mistakes to generate new questions \cite{wang2019upgrade, wang2021seeing}, using teacher-AI collaborative approaches where teachers receive AI suggestions\cite{lu2023readingquizmaker}, and fully automated techniques leveraging AI \cite{alsubait2016ontology, kurdi2020systematic, Das2021AutomaticQG, 10.1145/3231644.3231654, Majumder2014AutomaticSO, Majumder2015ASF, stasaski2017multiple, wang2022towards, ihantola2010review}. With the advancement of generative AI, there is a growing interest in using generative AI tools such as ChatGPT \cite{chatgpt} to create quiz questions. A handful of universities provide example prompts for instructors to create low-stakes assessment questions with ChatGPT \cite{miami, ucsd}. This movement has increased our likelihood of getting a large question pool, but how do we know whether the questions generated are of high quality?

In addition to face evaluation by experts and student peers \cite{lu2023readingquizmaker, wang2022towards, singh2021s}, psychometric methods are still mainstream to evaluate question item quality. 
Common psychometric methods evaluate test reliability by the internal consistency of question items within a test, e.g., using a Rasch model \cite{wilson2004rasch}, Item Response Theory (IRT) model \cite{harris1989irt}, or Cronbach's alpha \cite{cronbach1951coefficient, wang2019upgrade}. A unique challenge is that it requires substantial response data for such models to effectively prune out low quality (inconsistent) question items, making psychometric methods expensive and impractical to use in most college classrooms. Although teachers might be able to apply psychometric methods between semesters, most teachers do not have access to student response data the first time they assign the questions. 

We propose a modular prompt architecture \textit{Generative Students}, in which we leverage large language models (LLM) to simulate student profiles. In this paper, we demonstrate that we can have the generative students answer multiple-choice questions and use the responses to identify unreliable question items. The design of the prompt architecture is based on the Knowledge-Learning-Instruction framework \cite{koedinger2012knowledge}, which uses Knowledge Components (KCs) to define the elements students are expected to learn. In Generative Students, we simulate student profiles by the KCs they have mastered. In particular, for every given KC, the student may have mastered it, have confusion about it, or have not shown understanding of it. A student profile is essentially a function of the list of KCs they have mastered, have confusion about, or have not shown understanding of. We propose Generative Students as an approach that does not require students' historical performance data. Instead, we rely on instructors to provide input on the knowledge components required for skill mastery and the common misconceptions that they anticipate. This makes Generative Students potentially more generalizable to domains that do not have a lot of historical data.
Generative students can be created within seconds and produce a large amount of response data to a given set of questions. We aim to address the research questions of: 1) Is it possible to use LLMs to successfully simulate student profiles and generate believable answers to questions? 2) How do the generative students' responses contrast with authentic students' answers? 

We study these questions in the context of teaching and learning heuristic evaluation, a usability inspection method. The reason we pick heuristic evaluation as the subject domain is two-fold: 1) We have collected an authentic student response dataset with 20 multiple-choice questions (MCQs) on this topic, which enables the comparison between generative students and real students. 2) The topic of heuristic evaluation has well-defined knowledge components (KCs). In particular, learners need to master 10 Nielsen's heuristic rules and check them against a design. We can conveniently denote each of the 10 heuristics as a KC. 

Using the prompt architecture, we created 45 generative students with different mastery levels on the 10 KCs, and had them answer 20 MCQs. Each response to an MCQ is an API call to GPT-4. The LLM's response contains both the answer and a rationale for picking the answer. We first performed a qualitative analysis of the LLMs' responses showing that the generative students produced logical and believable responses that are aligned with their profiles. 

We then compared the 45 generative students and 100 real college students on their responses to the same set of 20 MCQs. To investigate how well a brutal force simulation approach would perform,
we added a third condition, where we simulated 45 students using random number generation. Each random student has a 70\% chance of getting each question correctly. The comparison between real students, generative students, and random students shows that, the real students and generative students have a high consistency in their responses measured by Pearson's correlation (r=0.72). However, the real students' and random students' answers are not correlated (r=-0.16). Moreover, we see a reasonable overlap on the easy and hard questions identified by generative students and real students, which shows potential of using generative students to signal questions that need revision. 


This study generates insights on creating LLM agents that have specific knowledge deficiency, when the LLM itself has perfect content knowledge. Specifically, we asked the LLM agents to play the role of a teacher and predict a student's answer to a question. 
This is the first study to our knowledge that shows promising results of leveraging LLM-simulated student profiles to help evaluate multiple-choice question items, without the requirement of student historical performance data. This opens up avenues for using generative students to support rapid prototyping and iteration of questions. We discuss the potential risks of the approach and the necessity of eliciting instructor (expert) input to steer the process.

\section{Related Work}
\subsection{Automatic Question Generation for Educational Purposes}

It has been a long standing interest of the Learning@Scale and AI in Education communities to study question generation techniques. 
One line of work uses crowdsourcing techniques \cite{singh2021s}. For example, UpGrade creates questions based on prior student solutions \cite{wang2019upgrade} and QMAps encourages students to generate questions for each other \cite{10.1145/3313831.3376882}. 
Another line of work develops end-to-end NLP models for question creation, which are good at creating factual questions ~\cite{Das2021AutomaticQG, kurdi2020systematic}, while not being able to generate questions that target higher Bloom goals \cite{bloom1965bloom}. 
On multiple-choice question (MCQ) generation, prior approaches used name entity recognition and topic modeling to identify salient sentences and extract keywords for question options ~\cite{Majumder2014AutomaticSO, Majumder2015ASF}. 
Recent work has also explored human-AI collaborative methods for MCQ creation, where instructors select text input for the options \cite{lu2023readingquizmaker}. Existing AI-assisted question generation systems face a common challenge, i.e., how to evaluate the quality of generated question items. In this work, we explore the feasibility of leveraging LLMs to simulate student responses and use them to evaluate auto-generated question items.

\subsection{Metrics and Approaches to Evaluate Questions}

Prior work has explored various strategies to evaluate question quality based on student data. Both learner subjective ratings \cite{Williams2016axis} and student performance data \cite{khairani2016assessing, etstoeflreport2011} were used to select high-quality content.
Item difficulty and discrimination indices evaluate whether the questions are differentiable \cite{khairani2016assessing}, while psychometric methods are used to evaluate the inner consistency of the questions \cite{cronbach1951coefficient, khairani2016assessing, wilson2004rasch, harris1989irt}. 
However, although these post-hoc analyses may favor future students, the question items with low quality in the first place might waste students' learning opportunities. 
Another line of approach focuses on evaluating questions based on the descriptions solely, using rubrics and guidelines like Bloom’s Taxonomy \cite{10.1007/978-3-030-78292-4_35} and item-writing flaws \cite{moore2023crowdsourcing}. 
The questions are viewed as lower quality if they only involve a low level of cognitive process \cite{costa2018evaluation}, and if they violate multiple rules in the item-writing flaws \cite{moore2023assessing}. Based on these rules, prior work explored automatic quality control approaches using supervised learning \cite{10.1007/978-3-030-78292-4_35}, neural networks \cite{ruseti2018predicting} and LLMs \cite{moore2023assessing} to reduce the need for human labor. However, these rule-based evaluations do not take into account students' obstacles in learning and could be biased by expert blindspots \cite{wang2021seeing}.


\subsection{Generative agents}
The success of LLMs in reasoning \cite{dasgupta2022language} and problem-solving abilities \cite{orru2023human} have attracted growing interest in LLM-powered generative agents \cite{wang2023survey, park2023generative}. Prior work has shown LLMs can be prompted to generate believable behavior \cite{park2023generative}, and even act like humans from certain sub-populations \cite{addlesee2023multi, markel2023gpteach}. By creating modules to simulate human memory, planning and reflection, the generative agents can mimic the logic in human behavior and result in a believable decision-making process \cite{park2023generative}. Prior work has shown the potential of generative agents in realistically simulating human behavior in various areas, including strategic gaming \cite{wang2023voyager, xu2023exploring}, social networking \cite{park2023generative}, and role-playing \cite{wang2023rolellm}. However, existing work focuses on simulating characteristics with different perspectives like occupations, personalities, values and relationships \cite{zhou2023sotopia}, where the character will make decisions to the best of their knowledge or memory. However, in this work, we aim to simulate agents that have knowledge deficiencies and will make mistakes when solving educational problems.

Recent work explored using generative agents in education \cite{han2023cheddar, markel2023gpteach, shaikh2023rehearsal, jin2024teach}. Xu and Zhang showed the potential of using LLM to model students' learning based on past assessment scores \cite{xu2023leveraging}. Researchers have also developed teachable agents, which provide practice opportunities for learners to identify knowledge gaps \cite{jin2024teach} and for instructors to receive feedback from students \cite{markel2023gpteach}. However, previous simulations often rely on historical student performance data. Our work presents a prompt architecture to create generative students using a data-sparse approach, where we rely on experts to provide a list of knowledge components required for skill mastery and a list of common student misconceptions. 

\subsection{Prompt Engineering}
Prompting has become a main way of utilizing and steering LLMs \cite{brown2020language}. Prior work has looked into different guidelines for creating effective prompts, including prompt structures \cite{brown2020language}, languages \cite{wei2022chain} and vocabulary \cite{oppenlaender2023prompting}. One direction of prompting is to use few-shot learning, where the task is demonstrated with several example input-output pairs \cite{brown2020language}. Wei et al. proposed Chain-of-Thought as a method to improve few-shot learning performance by including the thought process in the prompt \cite{wei2022chain}. However, prompt engineering is still a non-intuitive skill \cite{oppenlaender2023prompting} and can be challenging for non-experts. In our work, we investigate methods to prompt LLMs to behave like students with knowledge deficiencies. We summarize prompt engineering strategies using LLMs to simulate students.


\section{Generative Students Prompt Architecture}
We propose a prompt architecture based on a widely adopted framework in learning sciences, the Knowledge-Learning-Instruction (KLI) framework \cite{koedinger2015learning}. In particular, the KLI framework defines the finest units of information learners are expected to learn as Knowledge Components (KCs). KCs are supposed to be mutually exclusive and provide the basis to design instructional activities. 
Students acquire the KCs in their learning processes. The prompt architecture we propose defines a student profile as a function of the list of KCs the student has mastered, has confusion about, or has not shown evidence of knowledge on. 

\subsection{Implementation of Generative Students on Heuristic Evaluation}

In this paper, we implemented the concept of generative students on the topic of heuristic evaluation. Heuristic evaluation is a widely used usability inspection method, in which designers use rules of thumb to inspect the usability of user interfaces and identify design problems. We choose the topic of heuristic evaluation as our subject domain for two reasons. First, we have collected a student response dataset that contains 100 students' responses to 20 MCQs on heuristic evaluation. This makes it possible to compare generative students' responses to real students' responses. Second, the topic of heuristic evaluation has well-defined KCs, namely the 10 Nielsen's heuristic rules, as shown in Table~\ref{table:heurstics}. It removes the requirement for performing a cognitive task analysis in order to infer the KCs needed to complete this task. We acknowledge that the task of heuristic evaluation is unique in that the 10 KCs are clear-cut and concurrent, and that there are fewer dependencies among the KCs. We will discuss in later sections on the potential of generalizing Generative Students to other domains. 



\begin{table}[]
\caption{The topic of heuristic evaluation contains 10 knowledge components (KCs), namely the 10 Nielsen's heuristic rules. A generative student profile is a function of which KCs the student has mastered, has confusion about, or has shown no evidence of knowledge on.}
\begin{tabular}{|l|l|}
\hline
H1  & Visibility of system status                                                                         \\ \hline
H2  & Match between the system and the real world                                                        \\ \hline
H3  & User control and freedom                                                                           \\ \hline
H4  & Consistency and standards                                                                          \\ \hline
H5  & Error prevention                                                                                   \\ \hline
H6  & Recognition Rather than Recall                                                                     \\ \hline
H7  & Flexibility and Efficiency of Use                                                                  \\ \hline
H8  & Aesthetic and minimalist design                                                                    \\ \hline
H9  & Help users recognize, diagnose, and recover from errors 
    \\ \hline
H10 & Help and Documentation  \\  \hline     
\end{tabular}
\label{table:heurstics}
\vspace{-1pc}
\end{table}

\begin{figure*}
    \centering
    \includegraphics[width=\textwidth]{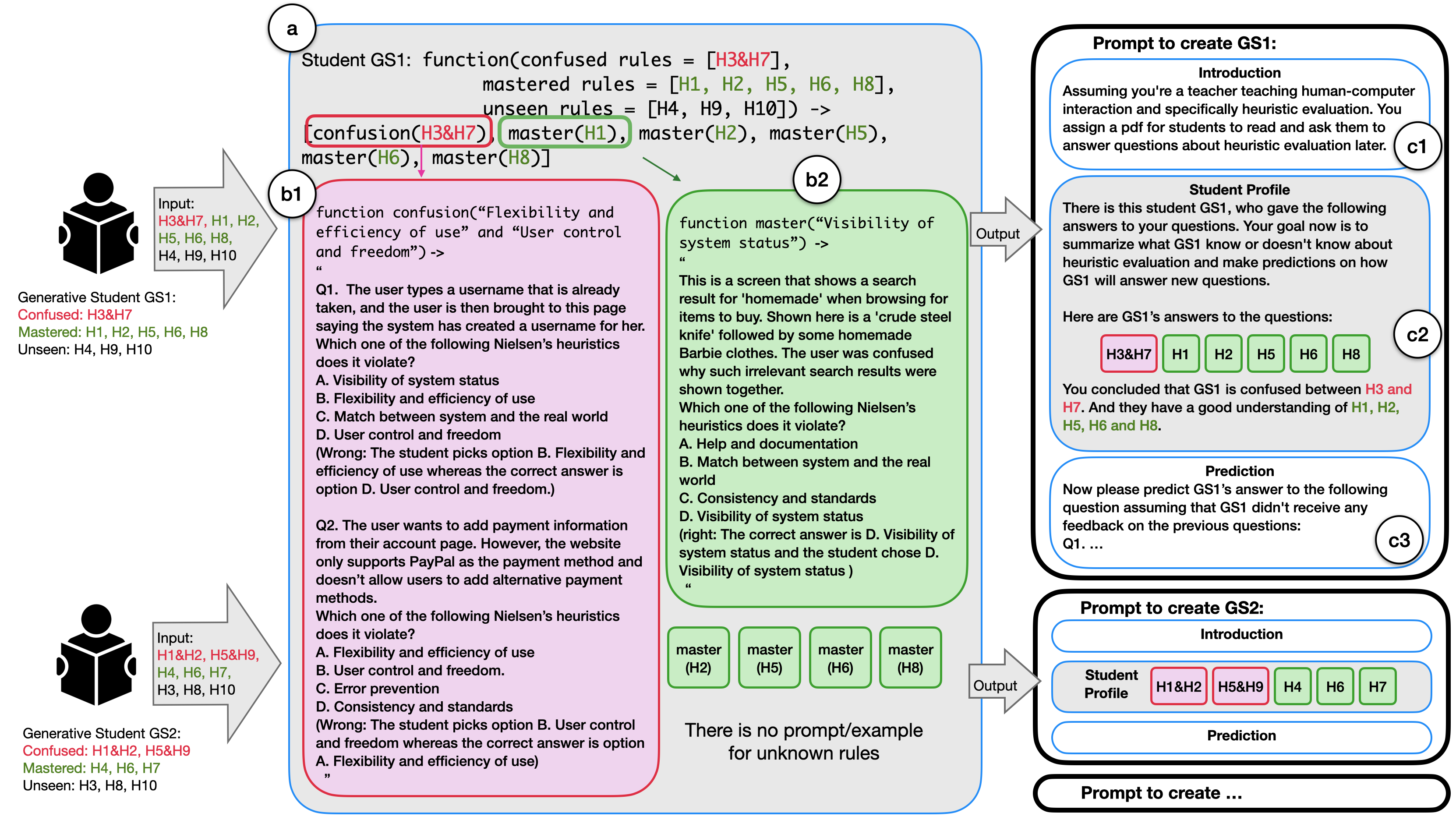}
    \caption{The prompt template has three main parts: 1) an introduction of the task (c1); 2) an illustration of the generative student profile (c2); 3) a new MCQ to which the generative student will answer (c3). The generative student profile is a function of the list of heuristic rules the student has mastered, has confusion about, or has no evidence of knowledge of (a). For each mastered heuristic rule, we used an example MCQ to indicate the student has sufficient knowledge (b2); For each pair of confusion heuristic rules, we used two example MCQs to indicate the student may mistakenly choose one over the other (b1).  
}
    \label{fig:prompt}
\vspace{-1pc}
\end{figure*}

\subsection{Final Prompt Structure and Examples} \label{sec:prompt_creation}



The template of the prompt we used to create generative students is shown in Figure \ref{fig:prompt}. The prompt has three main parts: 1) an introduction of the task; 2) an illustration of the generative student profile; 3) an MCQ to which the generative student will produce an answer.
First, the introduction specifies that the model is playing the role of a teacher predicting a student's answer. We will explain the rationale of asking the model to simulate a teacher instead of directly simulating a student answering a question in \ref{sec:roleplay}. Second, the generative student profile is a function of lists of heuristic rules the student has mastered, has confusion about, or has shown no evidence of knowledge of. For example, generative student 1 (GS1) has mastered five rules, is confused between two rules, and has not shown evidence of knowledge of three rules. For each mastered rule, it calls the mastered rule prompt function, which gives an example question demonstrating the student has given a correct answer to this question. The confusion prompt requires two rules as the input, and uses two example questions to show that the student has confusion between these two rules. There is no prompt for the unknown rules. Third, the model is asked to predict an answer to a new MCQ. With a different input on the list of mastered, confusion, and unknown rules, the template generates a different generative student profile (e.g., GS2 in Figure~\ref{fig:prompt}). 

\subsection{Input to the Prompt Template}
The example questions and answers used in the prompt are provided by an instructor who has been teaching this topic for 5 years. The instructor provided one example question for each of the 10 heuristic rules, indicating the student has mastered the rule. Moreover, the instructor suggested 2 pairs of common confusions: between "H3-User control and freedom" and "H7-Flexibility and efficiency of use", and between "H5-Error prevention" and "H9-Help users recognize, diagnose, and recover from errors". 
We also created 2 pairs of random confusions. 
This gives us 4 pairs of confusion rules. For each pair, the instructor provided us with 2 example questions, where both heuristic rules are in the options and the correct answer is one of them, as shown in Figure~\ref{fig:prompt}(b1). Please note that the example questions the instructor provided to us were similar in style to the 20 new questions we're producing answers on. But the 20 questions are completely new questions that the generative students have not answered before. To summarize, the Generative Students prompt architecture requires the following input from an expert:
\begin{itemize}
    \item Knowledge components(KCs) required to perform a task.
    \item An example question per KC with the correct answer to demonstrate a student has mastered this KC. 
    \item Common student misconceptions (optional). In the case of heuristic evaluation, a misconception involves two KCs. 
    \item Example question(s) per confusion with an incorrect answer to demonstrate a student has confusion about this KC. In the case of heuristic evaluation, it requires two example questions. 
\end{itemize}

\subsection{Takeaways from the Prompt Engineering Process}
In this section, we describe our takeaways from the prompt engineering process that led to the final prompt as shown in Figure~\ref{fig:prompt}. 

\subsubsection{Providing example MCQs and answers improves performance.}
We found that using example questions to indicate the student has mastered or has confusion about a rule is more effective than simply stating it, in line with prior work \cite{pmlr-v202-shi23a, wei2022chain}. In particular, for the confusion prompt component, we first tried specifying that the student was confused about one rule. However, it will bias the model towards always picking or not picking one choice. From our trial-and-error process, the current prompt that takes two rules as arguments works best in simulating a student's confusion. Moreover, we need two example questions to demonstrate the student can make a mistake in both directions. When we only use one example question in the prompt, the model would mistakenly think the generative student will always pick one rule over the other.

\subsubsection{Asking the model to role-play as an instructor and predict the generative student's answer helps.}\label{sec:roleplay}
Instead of prompting the LLM to act as a student and "answer" the questions directly, we ask it to act as a teacher who wants to "predict" the student's answer. 
We found that when asked to answer the questions based on the student's profile, the LLM is more likely to answer based on its prior knowledge. For example, even when we specify in the prompt that the student has confusion about a rule, the model will still answer a related question correctly. On the other hand, when we prompt the model to act as a teacher to predict the student's answer, the model's performance is better aligned with the student's profile.


\subsubsection{Using unknown rules to increase uncertainty in the predicted answers}
To better simulate real students' responses to the questions, we'd like to introduce some uncertainty on the generative students' answers. We found that specifying some unknown rules, i.e., leaving them blank without any explicit prompting, achieves this goal. For example, for GS1, H4, H9, and H10 are unknown rules. We do not have any prompt components that specify the student's knowledge on these KCs. It turns out to be effective in introducing uncertainty in the generative students' answers. 


\subsubsection{Introducing uncertainty within the confusion prompt component by providing both positive and negative examples.}
In reality, even when a student has confusion between two rules, they may still get easy questions correct. To simulate this uncertainty, we created a variation of the confusion prompt component, where the student has shown some understanding, but hasn't mastered it yet. 
We introduce example questions with varying difficulty in the prompt. For example, we specify that the generative student can answer the easy questions correctly while making errors on more difficult questions. 
We found that the variation of the confusion prompt introduced more uncertainty that aligned with the prompt specification.

\subsubsection{Using the prompt to get a generative student's response to one question at a time gives better results.} \label{sec:predict_single_question}
We found that prompting the generative students to answer 20 questions all at once did not work well. First, due to the token limit in each response, the response to each question is shorter. As a result, we observe shallower reasoning. Second, if asked to predict 20 questions at a time, the model will use its answer to a former question to predict a later question. Moreover, a previous question's response may also override the generative student's profile, leading to answers misaligned with the profile.

\section{Generative Students Response Dataset}

\subsection{Creation of 45 Generative Students}

As we noted earlier, each generative student is a function of the list of rules they have mastered, have confusion about, or have no evidence of knowledge of. We can pass the list of rules as parameters to automatically create the generative students, as shown in Figure~\ref{fig:prompt}. A decision we need to make is how knowledgeable the generative student is, e.g., the student can master 3 rules, 5 rules or 9 rules. 
In this experiment, we created a suite of 10 struggling students, 30 average students, and 5 advanced students, as shown in Table~\ref{table:ruledistribution}. The advanced students are more knowledgeable because they have less confusion as specified in the prompts. 
With the distribution set (as shown in Table~\ref{table:ruledistribution}), one can randomly select heuristic rules from the list to automatically create generative students. 
In this study, we applied a semi-random approach instead of a fully random one to better contrast different generative student profiles. 
We created pairs of similar student profiles, where only one variable differed with everything else being the same between the two profiles. 
For example, to create two generative students with the same set of mastered rules, but a different pair of confused rules, we would randomly pick a confusion pair for each student, and then randomly pick the same set of 5 mastered rules for both of them.

\begin{table}[]
\caption{The distribution of the number of mastered, confused and unknown rules among the 45 generative students. }

\begin{tabular}{|p{1.8cm}|p{2cm}|p{1.6cm}|p{1.8cm}|}
\hline
\# of mastered HE rules & pair of confused HE rules                                           & \# of unknown rules & \# of generative students \\\hline
5                              & 2                                                                          & 3                  & 15 (Average)                       \\\hline
7                              & 2                                                                          & 1                  & 15  (Average)                      \\\hline
5                              & 4                                                                          & 1                  & 5    (Struggle)                     \\\hline
3                              & 4                                                                          & 3                  & 5     (Struggle)                    \\\hline
5                              & 2 (focused confusion prompt) & 3                  & 5 (Advanced)             
\\
\hline
\end{tabular}
\label{table:ruledistribution}
\vspace{-1pc}
\end{table}

\subsection{Implementation}
We used the OpenAI API to provide inputs and get responses from GPT-4. As shown in Section \ref{sec:predict_single_question}, each API call contains only one generative student profile, and only predicts the response to one question. To ensure getting the most reliable prediction, we set the temperature of the model to 0, which ensures that we are getting the exactly same result from GPT with the same prompt.


\subsection{Example GPT-4 Responses}

With each API call, the model will output a predicted response, i.e., which option the student would pick, and a rationale explaining why the model believes the student will pick this option. Here we present two examples of the responses generated by GPT-4. In this example, the generative student has confusion between "Flexibility and efficiency of use" and "User control and freedom". The correct answer to the question is "User control and freedom", and "Flexibility and efficiency of use" is among the options. It is predicted that John would incorrectly choose "Flexibility and efficiency of use" based on the reasoning: 
\begin{quote}
    Given that John incorrectly chose "Flexibility and efficiency of use" over "User control and freedom" for a question where the latter was the correct answer, there's a possibility that he might repeat this mistake if he perceives the new scenario as a limitation in the system's flexibility or efficiency, rather than an issue of user control.
\end{quote}

In another question, the correct answer is a rule John has mastered, "Visibility of system status". None of John's confused rules are among the options. It is predicted that John would choose the correct answer with the reasoning:
\begin{quote}
    Considering John's track record, it's plausible that he could again accurately recognize this scenario as a violation of "Visibility of system status." His confusion doesn't directly apply to this scenario, so it's less likely to influence his answer here.
\end{quote}

\section{Evaluation of the Outputs by Generative Students}

\subsection{Methods}
To investigate whether the answers are aligned with the generative students' profiles, we applied both quantitative and qualitative analyses. First, we counted the number of instances the generative students answered correctly across different conditions, e.g., when the student has mastered a rule versus not. Second, 
we analyzed the rationale the model generated using affinity diagrams \cite{moggridge2007designing}, where we iteratively grouped the reasoning and identified common themes.

\subsection{Generative Students' Answers are Generally Aligned with the Student Profiles}


\begin{table}[htbp]
\caption{Performance by the generative students who have mastered 5 rules, have confusion between 1 pair of rules. In the target question, when the correct answer is a mastered rule, students demonstrate good performance. When the correct answer is one rule in the confused pair, the accuracy is generally low. When the correct answer is an unknown rule, there is a 30\%-50\% chance that the student can answer correctly. In all three conditions, students' performance is lower when the confused rules are present in the distractors.}
\begin{tabular}{|l|l|l|l|l|}
\hline
\begin{tabular}[c]{@{}l@{}}Correct \\Answer\end{tabular} & \begin{tabular}[c]{@{}l@{}}Confusion \\ in Other \\ Options\end{tabular} & \%Correct & \begin{tabular}[c]{@{}l@{}}Total\\ Responses\end{tabular} & \begin{tabular}[c]{@{}l@{}}\% Choosing \\ Confused \\Rule\\\end{tabular} \\ \hline
Mastered       & No                                                                       & 85.2      & 64                                                        &    -                                                                                   \\ \hline
Mastered       & Yes                                                                      & 72.4      & 125                                                       & 59.4                                                                                  \\ \hline
Confused      & No                                                                       & 35.6      & 52                                                        &    -                                                                                   \\ \hline
Confused      & Yes                                                                      & 11.0      & 41                                                        & 82.2                                                                                  \\ \hline
Unknown         & No                                                                       & 52.1      & 47                                                        &    -                                                                                   \\ \hline
Unknown         & Yes                                                                      & 34.5      & 71                                                        & 46.2                                                                                  \\ \hline
\end{tabular}
\label{table:5-3-2}
\vspace{-0.5pc}
\end{table}

\subsubsection{Generative students are likely to answer a question correctly when the correct answer is a "mastered" rule.} \label{sec:mastered_rule}
When the correct answer of the MCQ is a mastered rule, and no confused rules are present in the distractors, the student is likely to answer correctly ($85.2\%$ of the time). 
For example, GS9 has mastered the rule "User control and freedom", he is predicted to correctly answer a question on the same heuristic because \textit{"His past performance indicates a good grasp of this particular heuristic, suggesting he is likely to apply it correctly again."}
When confused rules present in the options, the accuracy is slightly lower ($72.4\%$). Here is one sample of GS8 predicted to answer Q5 correctly where the correct answer is a mastered rule and both confused rules are among the options. GPT-4 reasoned that \textit{"the new question directly concerns the visibility of system status, an area John previously demonstrated an understanding of. Additionally, the new question does not directly involve distinguishing between 'Error prevention' and 'Help users recognize, diagnose, and recover from errors,' areas where John showed confusion."}


We summarized two scenarios where the student answers the question wrongly when the options contained a confused rule ($27.6\%$). First, students might incorrectly choose the confused rules. 
Second, the generative student may pick a suboptimal answer but the reasoning shows a legitimate understanding of the mastered rule. This happened when the question stem was vague and multiple answers could be correct. 
For example, Q20 is about the system not supporting using the 'tab' key to navigate the form, the correct answer is "Flexibility and efficiency of use". However, given the prevalence of the 'tab' key function, this scenario can also be interpreted as violating "Consistency and standards". Although GS9 has mastered "Flexibility and efficiency of use", they are predicted to choose "Consistency and standards" for this reason.

\subsubsection{Generative students are likely to answer the questions wrong when the correct answer is a "confused" rule.} \label{sec:confusing}

When the correct answer is a heuristic that the student has shown confusion about, the student will have a high likelihood of getting it wrong. The chance of getting it wrong is in particular high, when the other rule in the confusion pair is present in the options, as shown in Table~\ref{table:5-3-2}.



\subsubsection{Generative students could correctly answer a question when the correct answer is a "confused" rule, if the other rule in the confusion pair is not among the options.}

For example, the correct answer for Q9 is "Help users recognize, diagnose, and recover from errors". Although GS8 is confused between "Error prevention" and "Help users recognize, diagnose, and recover from errors", they are predicted to answer this question correctly since "Error prevention" is not in the options. GPT-4 predicts
\textit{"it's more likely that they might choose the option that is closest to dealing with errors ... 'Help users recognize, diagnose, and recover from errors'"}. The generative student might also get correct by eliminating other options. 
For example, GS2 confuses about the correct answer in Q6, but since the other confusion rules are not among the options, and they have mastered the rules in all the other options, GPT-4 predicts that \textit{"he might eliminate these options because (they could identify) they (other options) don't fit the scenario as well as 'Visibility of system status' does."}

\subsubsection{Generative students are likely to be wrong when the correct answer is an "unknown" rule, and there is a "confused" rule in the option.}
However, when the options do not contain a confused rule, the generative students show a slightly above $50\%$ chance to get the questions right. For example, Q2 doesn't contain any confused rules for GS3 in the options, and the correct answer is an unknown rule. GS3 is predicted to answer it correctly because \textit{"the clear match between the scenario's description and the fundamental concept of 'User control and freedom'."} and the fact that \textit{"the situation doesn't directly involve error messages or prevention"}.
Moreover, GPT-4 also predicts the student's knowledge of the unknown rule based on their knowledge of related heuristic rules. 
For example, the correct answer to Q4-"Consistency and standards"- is an unknown rule to GS8. Q4's options do not contain any confused rules for GS8. GPT-4 predicts that since they \textit{"correctly answered questions related to user interface design and user experience consistency ('Visibility of system status,' 'Aesthetic and minimalist design', and 'Match between system and the real world,' )", "there is a reasonable chance that they might select the correct answer"}. 

\subsection{Controlled Variable Generation in the Prompt Leads to Probable Outcomes}
As mentioned earlier, we applied a semi-random approach when selecting the list of heuristic rules to create the generative students. We created pairs of similar student profiles, where only one variable differed with everything else being the same between the two profiles. This allows us to examine whether the change of one variable in the prompt leads to outputs that align with our expectations. 

\subsubsection{Changing only the confusion pair in the student profile yields reasonable outputs.}
We created multiple pairs of similar student profiles where the only difference was the confusion pair. We found that the generative students would produce outputs that are aligned with our expectations. 
For example, the correct answer for Q15 is "Flexibility and efficiency of use". GS4 has shown confusion between "Flexibility and efficiency of use" and "Recognition rather than recall", so they are predicted to choose "Recognition rather than recall" in this question. 
On the other hand, GS11 has the same set of mastered rules as GS4, and the only difference between their profile is that GS11 is confused between "Flexibility and efficiency of use" and "User control and freedom". For this question, GS11 is thus predicted to choose "User control and freedom".

\begin{figure*}
    \centering
    \includegraphics[width=\textwidth]{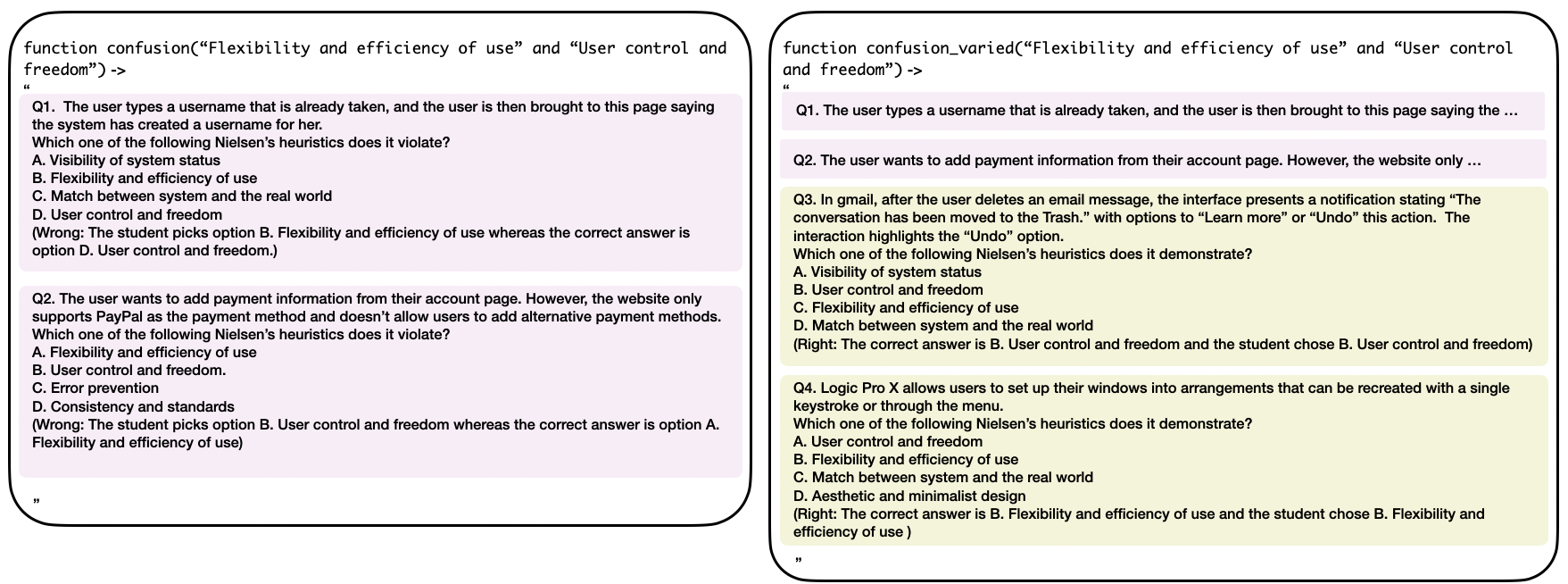}
    \caption{The focused confusion prompt (right) contains the two original questions that the student got wrong (Q1, Q2), and two additional examples to show that the students may answer the easy questions correctly (Q3, Q4). Generative students who use the focused confusion prompt are expected to have better overall performance. The focused confusion prompt aims to introduce more uncertainty to better simulate realistic scenarios.}
    \label{fig:confusion-varied}
\vspace{-1pc}
\end{figure*}

\subsubsection{A more knowledgeable student is more likely to answer questions on unknown rules correctly}


Consider a generative student that has mastered 5 rules, has 1 pair of confused rules, and 3 unknown rules. If we make this student more knowledgeable by adding 2 mastered rules, while keeping the original confusion pair, our experiment shows that the more knowledgeable student will be more likely to correctly answer questions on unknown rules. This aligns with our expectations.
For example, Q10's correct answer "Visibility of system status" is a confused rule to GS2. Among the other 3 options of Q10, GS2 has mastered one rule and the other two rules are unknown. GS2 is predicted to answer Q10 incorrectly. When we create a new generative student GS12 with additional mastery of the two unknown options, they answer Q10 correctly, since they can successfully eliminate the options. 
Moreover, an increased understanding of heuristic rules in general indicates an improved overall ability to identify heuristic rules. In GS12's case, the increased understanding in other heuristics indicated that they might be correct in answering questions with unknown rules as reasoned by the model: \textit{"[because] their overall good performance on questions related to the usability heuristics that directly impact user interaction and control."}

\subsection{The Focused Confusion Prompt Introduces Uncertainty and Improves Students' Overall Performance on Questions Related to the Confused Rules}
\begin{table}[htbp]
\vspace{-0.5pc}
\caption{We contrast the performance of generative students who used the original confusion prompt versus the focused confusion prompt. The focused confusion prompt suggests that the student will answer easy questions related to the confusion correctly. It aligned with our expectation that students using the focused confusion prompts have better performance.}
\begin{tabular}{|p{1.3cm}|p{1.3cm}|p{1.1cm}|p{1.3cm}|p{1.4cm}|}
\hline
\begin{tabular}[c]{@{}l@{}}Correct \\ Answer\end{tabular}       & \begin{tabular}[c]{@{}l@{}}Confusion \\ in Other \\ Options\end{tabular} & \%Correct & \begin{tabular}[c]{@{}l@{}}Total\\ Responses\end{tabular} & \begin{tabular}[c]{@{}l@{}}\%Choosing\\Confused\\Rule\end{tabular} \\ \hline
Confused                                                      & No                                                                       & 26.9      & 39                                                        &   -                                                                                    \\ \hline
Confused                                                      & Yes                                                                      & 0         & 30                                                        & 25                                                                                    \\ \hline
\begin{tabular}[c]{@{}l@{}}Confused\\ V-Prompt\end{tabular} & No                                                                       & 61.5      & 13                                                        &     -                                                                                  \\ \hline
\begin{tabular}[c]{@{}l@{}}Confused\\ V-Prompt\end{tabular} & Yes                                                                      & 40.9      & 11                                                        & 100                                                                                   \\ \hline
\end{tabular}
\label{table:confused-rule}
\vspace{-1pc}
\end{table}
We implemented a focused confusion prompt to indicate that even when the student is confused between two rules, there is a chance that they may answer easy questions on the confused rules correctly. The GPT-4 output is aligned with the expectation, as shown in Table~\ref{table:confused-rule}. The student profiles with the focused confusion prompt have a higher likelihood of answering the questions correctly. 
We compare two generative students, GS1 and GS21, who have the same profile, except that GS21 uses the focused confusion prompt on the same pair of confused rules. 
GS21 is predicted to answer Q14 correctly, where the correct answer is a confused rule for both GS21 and GS1, whereas GS1 is predicted to answer it incorrectly. The reasoning for GS21's correct answer is that 
\textit{"considering that they correctly identified 'Flexibility and efficiency of use' in a previous question where it was indeed the correct answer, there's a good chance he will choose the correct answer this time."}

\section{Comparison Between Real Students, Generative Students, and Random Students}


\subsection{Datasets}
\subsubsection{Real Students' Response Dataset}

The same set of the 20 MCQs have been previously assigned in a college-level course at an R1 institution in 2021 as a homework assignment. We got IRB approval to collect student responses from that class. Students in the course were asked to complete the assignment through a website that contains the same 20 MCQs on the topic of heuristic evaluation. 100 students completed the assignment. 

\subsubsection{Random Students' Response Dataset}
To investigate how well random simulations perform on this task and compare our principled simulation approach with a random one, we designed a baseline condition where student responses were generated randomly. We refer to these as random students. 
The random students are created based on random number generation. For each question, there is a 70\% chance of getting the question correctly. We generated 45 random students.

\subsection{Methods}

First, to check the consistency of real students' responses with generative and random students' respectively, we computed Pearson's correlation using the students' average score on each question. 
Second, we used Cronbach's Alpha to measure the internal consistency of each dataset. Third, we identify hard and easy questions based on the responses. We employed two thresholds: if the average score is above 80\%, it's considered to be an easy question, and if the average score is below 40\%, it's considered to be a hard question. We compared across the three conditions to assess the overlap on the easy and hard questions identified. Moreover, for the questions where real students and generative students yielded different results, we performed an error analysis analyzing the distribution of the options students picked.



\subsection{Results}

First, as shown in Table~\ref{table:average}, generative students' responses show a high correlation with real students' responses, with a Pearson's correlation of 0.72. On the other hand, the correlation between the random students and the real students is only -0.16. 

Second, the generative students' responses dataset shows high internal consistency as measured by Cronbach's Alpha ($0.6176$), also shown in Table~\ref{table:average}. The internal consistency is comparable to that of the real students' response dataset ($0.559$). However, the random students' response data has low internal consistency ($0.042$).


\begin{table}[htbp]
\caption{Average scores for each of the 20 MCQs across the three response datasets. Green text indicates easy questions (>0.8) and red text indicates hard questions (<0.4). The questions that receive very low scores may suggest that the clarity of the questions needs to be improved.}
\resizebox{0.5\textwidth}{!}{%
\begin{tabular}
{|l|l|l|l|l|l|l|l|}
\hline
 & \begin{tabular}[c]{@{}l@{}}Real \\ Stu-\\dents\end{tabular} & \begin{tabular}[c]{@{}l@{}}Generative\\ Students\end{tabular} & \begin{tabular}[c]{@{}l@{}}Random\\ Stu-\\dents\end{tabular}& & \begin{tabular}[c]{@{}l@{}}Real \\ Stu-\\dents\end{tabular} & \begin{tabular}[c]{@{}l@{}}Generative\\ Students\end{tabular} & \begin{tabular}[c]{@{}l@{}}Random\\ Stu-\\dents\end{tabular}\\ \hline
Q1                                                                               & 0.76                                                     & 0.54                                                          & 0.73                                                         & Q11                                                                              & 0.84 (+)                            & 0.8 (+)                           & 0.6                                                       \\ \hline
Q2                                                                               & 0.69                                                     & 0.51                                                          & 0.67                                                         & Q12                                                                              & 0.88 (+)                             & 0.67                                                           & 0.76                                                         \\ \hline
Q3                                                                               & 0.56                                                     & 0.22 (-)                                & 0.73                                                  
& Q13                                                                              & 0.16 (-)                              & 0.04 (-)                                   & 0.69                                                  \\ \hline

Q4                                                                               & 0.33 (-)                              & 0.53                                                          & 0.69                                                         & Q14                                                                              & 0.72                                                     & 0.52                                                          & 0.73                                                         \\ \hline
Q5                                                                               & 0.74                                                     & 0.57                                                          & 0.56                                                       &
Q15                                                                              & 0.52                                                     & 0.26 (-)                                   & 0.82 (+)                           \\ \hline
Q6                                                                               & 0.79                                                     & 0.67                                                           & 0.76                                                         &
Q16                                                                              & 0.69                                                     & 0.34 (-)                                   & 0.71                                                         \\ \hline
Q7                                                                               & 0.84 (+)                              & 0.72                                                          & 0.62                                                         &
Q17                                                                              & 0.36 (-)                              & 0.34(-)                                   & 0.62                                                  \\ \hline
Q8                                                                               & 0.64                                                     & 0.67                                                           & 0.82 (+)                                  &
Q18                                                                              & 0.59                                                     & 0.8 (+)                                   & 0.71                                                         \\ \hline
Q9                                                                               & 0.45                                                     & 0.37 (-)                                   & 0.8 (+)                            & Q19                                                                              & 0.85(+)                              & 0.94 (+)                                   & 0.62 \\ \hline
Q10                                                                              & 0.57                                                     & 0.79                                   & 0.64                                                         &
Q20                                                                              & 0.37 (-)                              & 0.14 (-)                                   & 0.71                                                         \\ \hline

\multicolumn{5}{|l|}{Pearson’s R with real student averages} & 1                                                        & 0.72                                                          & -0.16                                                        \\ \hline
\end{tabular}
}
\label{table:average}
\vspace{-0.5pc}
\end{table}

\begin{table}[]
    \centering
    \caption{The real students' and the generative students' response datasets have a comparable medium-to-high value of Cronbach's Alpha indicating good internal consistency, in contrast with the random students' response dataset.}
    \begin{tabular}{|c|c|c|c|}
        \hline
        &\begin{tabular}[c]{@{}l@{}}Real \\ Students\end{tabular} & \begin{tabular}[c]{@{}l@{}}Generative\\ Students\end{tabular} & \begin{tabular}[c]{@{}l@{}}Random\\ Students\end{tabular} \\
        \hline
         Cronbach's Alpha & 0.559 & 0.6176 & 0.042\\
         \hline
    \end{tabular}
    \label{tab:my_label}
    \vspace{-0.5pc}
\end{table}

We see 3 overlaps between the generative and real students' datasets among the hard questions identified, and 2 overlaps among the easy questions identified. However, there does not exist any overlap between the real and the random student datasets.
The questions that received low scores may suggest that the clarity of the questions needs to be improved. For these hard questions, we further analyzed the distribution of students' answers on the options to reveal the sources of mistakes. 
In Table \ref{table:frequent_wrong}, we present the frequent wrong answers chosen by over $25\%$ of the students. There is considerable overlap in the wrong options students picked between the real student and generative student response datasets. If instructors are interested in improving the clarity of the questions, they could leverage such information.

\begin{table}[]
\caption{There is considerable overlap between the real students and generative students on the distracting options they picked (chosen by over $25\%$ of students) for the hard questions. Instructors might leverage such information to improve the clarity of the questions. }
\begin{tabular}{|l|l|l|l|l|l|l|}
\hline
                   & Q3 & Q9 & Q13 & Q15 & Q17 & Q20 \\ \hline
Student Dataset    & A  & A,~C & C,~D  & D   & A   & C.~D  \\ \hline
Generative Dataset & A  & A,~C & C,~D  & D   & B,~C  & C,~D  \\ \hline
\end{tabular}
\label{table:frequent_wrong}
\vspace{-1pc}
\end{table}



\subsection{Error Analysis}

We further performed an error analysis to shed light on what caused the generative students to answer questions differently from the real students.

\subsubsection{Generative students had better performance on some questions because the real students' confusion was not included in the profiles.}

The generative students show higher performance in Q4, Q10 and Q18 in comparison with the real students. 
One reason is that most of the options in these questions are not among the pairs of confusions. As a result, most generative students won't find any of the options to be confusing and will answer them correctly.  
However, real students show coherent confusion in these questions. 76\% (51 out of 67) of the real students who got Q4 wrong and 49\% of the students who got Q18 wrong chose "Flexibility and efficiency of use" instead of "Consistency and standards". This suggests that the inclusion of a more diverse set of confusion KCs may improve the proximity between the generative students' with real students' responses. 


\subsubsection{Students' confusion may be over-emphasized or over-generalized leading to more pessimistic predictions on some questions.}
The generative students show a higher tendency to fail questions when the options contain a heuristic they have confusion about. 
Although real students may also make repeated errors, the portion is lower. 
For example, about $25\%$ of the generative students lean to the wrong option "Visibility of system status" for Q7. The same trend also appears in the real students, but it only takes up $7\%$ of the responses. 

\subsubsection{LLM may lose focus on the question}
When the question description emphasizes a seemingly positive feature, the generative students may misinterpret the question as asking for what heuristic it describes, instead of what it violates. For example, the description of Q3 reads \textit{"There are several ways to browse different categories of products on the same page. The user can either click the 'Shop by category' dropdown menu or click on the tabs from the main page."}. Many of the generative students \textit{"interpret the presence of multiple ways to browse as a feature that enhances flexibility and efficiency"}.



\section{A Potential Use Case Leveraging Generative Students to Improve Question Quality}
We did a case study with an instructor who was teaching heuristic evaluation in the Spring semester of 2024 at an R1 institution. Specifically, we presented them with the original 20 questions, generative students' responses, and asked them to revise some questions based on the signals. The instructor specifically picked Q3, Q9, Q13, and Q20 which received average scores below 0.4 when answered by generative students (as shown in Table~\ref{table:average}. The instructor considered these 4 questions to be badly worded and improved the clarity based on the wrong options the generative students picked. We then ran a classroom experiment including the original and improved versions of the questions in a required quiz in a class with $280$ students. 
The case study and the classroom experiment are IRB approved. 

In the classroom experiment, all consented students needed to answer a 7-question quiz on heuristic evaluation. We created two versions of the quiz following a crossover design \cite{wang2021seeing}. Quiz version A contains the improved version of Q9 and Q20, with the original version of Q3, Q13, Q1, Q5, Q7. Quiz version B contains the improved version of Q3 and Q13, with the original version of Q9, Q20, Q1, Q5, Q7. All students were randomly assigned to a quiz version. Q1, Q5 and Q7 are shared between the two versions and used to control for the two groups' prior knowledge. 

\subsection{Real Students Got Better Performance on the Revised Questions}
A randomization check showed that students in the two quiz versions showed similar performance on the shared questions Q1, Q5 and Q7, indicating that the comparison between the two groups was fair.
To investigate whether the revised questions appeared to be less difficult as intended, we built a mixed-effects logistic regression model where the dependent variable is the question score (0 being incorrect and 1 being correct), and the fixed effect is the question form, i.e., whether the question is original or modified. To account for different question difficulty and varying student abilities, we included a random slope for each question, and a random intercept for each student and each question \cite{wang2021seeing}. We found that the revised questions led to a significant increase in the question score (z=-2.538, p=0.01 < 0.05). The average score improvement is $0.248$. The average score on each question is shown in Table \ref{table:result}.
In particular, there is no improved performance on Q3, probably due to the fact that students got reasonably high scores on the original question. 

\begin{table}[]
\caption{Students are randomly assigned to answer Version A or B of the quiz. For Q3, Q9, Q13 and Q20, the underlined version is the revised question leveraging generative students' signals (including the wrong options picked by generative students). Q1, Q5, and Q7 are baseline questions that are the same across the two versions. We observed that there is a significant improvement in the average question score between the original and the revised version. 
}
\vspace{-0.5pc}
\begin{tabular}{|l|l|l|l|l|l|l|l|}
\hline
        & Q3          & Q9          & Q13         & Q20         & Q5    & Q1    & Q7    \\ \hline
Version A (n=147) & 0.84       & {\ul 0.86} & 0.29       & {\ul 0.83} & 0.88 & 0.94 & 0.96 \\ \hline
Version B (n=133) & {\ul 0.81} & 0.68       & {\ul 0.80} & 0.50       & 0.81 & 0.93 & 0.92 \\ \hline
\end{tabular}
\label{table:result}
\vspace{-1pc}
\end{table}

\section{Discussion and Future Work}

Our work shows promises of using the Generative Students prompt architecture to simulate student profiles that can generate believable and logical responses to MCQs. One potential avenue of this work is to help instructors quickly evaluate an initial set of questions, identify bad items and improve them before assigning to real students. In this section, we discuss how far we are from that goal based on the results presented from this study. 

First, we see promising results on the high correlation between generative students' and real students' responses, and overlap on the questions that students answered poorly. We discussed several prompt engineering takeaways for simulating student behaviors. First, describing the task as a pedagogical prediction leads to predictions more aligned with the student profiles. 
Second, illustrating students' knowledge with example questions and answers result in predictions that better align with the profiles.
Third, we explored methods to introduce diversity and uncertainty when simulating students, including a focused confusion prompt, an unknown component in the prompt architecture, and including example questions of varying difficulty when specifying the student profiles. 
On the other hand, we also revealed different reasoning models behind real and generative students. For example, generative students might appear to be more stubborn as they repeatedly make similar mistakes. It requires more experiments to enhance the proximity of the generative students' responses. For example, future work could explore including a more diverse set of confused rules, and introducing more prompt variations similar to the focused confusion prompt to increase uncertainty. 

Second, we propose a general-purpose architecture for simulating student profiles.
Although we only demonstrated our pipeline on one topic, i.e., heuristic evaluation, the architecture may be applied to other domains. For topics where the KCs are less well-defined and have more dependencies, expert input and specifications are required to ensure the generative students produce reliable and believable outputs.
In future work, we plan to collaborate
with instructors to understand how they would define KCs, structure the prompts, gather examples, and interpret the results. We also aim to investigate the feasibility of this approach when the instructors are not able to articulate the KCs required for skill mastery. We also plan to understand the time commitment from instructors to ensure that this is a reasonable amount of effort for them to prototype and iterate on their questions. 

Third, the confusions we included in the student profiles are based on experts' understanding of novice students' challenges, which may not be comprehensive. We plan to further explore methods to communicate students' knowledge to LLMs. For example, students' historical performance data may provide a more accurate representation of student knowledge, however it requires a lot of data input and may be less scalable for everyday teaching. Future work may explore methods to combine the expert-guided approach as proposed in Generative Students with a small set of student performance data to improve the simulation output.

Fourth, the case study showed that generative students could provide signals for instructors to improve their questions, e.g., help them identify difficult questions, and give them insights on frequently picked wrong options. The case study showed that an instructor could indeed leverage such information to iterate on the questions and make them less difficult. However, we acknowledge that "less difficult" does not necessarily indicate higher educational value. It requires further careful analysis to understand whether and when "less difficult" is desirable for instructors and students.











\section{Conclusion}

We propose Generative Students, a prompt architecture using LLM to simulate student profiles and produce reliable and believable responses to MCQs. With knowledge components (KCs) identified for a given topic, a generative student profile is a function of the list of KCs the student has mastered, has confusion about, or has shown no evidence of knowledge of. We show that providing concrete question-and-answer examples and having the model play the role of a teacher to predict student performance helps simulate believable student behaviors. 
Our results suggest that generative students' responses to the MCQs are aligned with their profiles, and exhibit a strong correlation with those of actual students. A subsequent case study demonstrates that generative students provide useful signals for an instructor to identify badly-worded MCQs and improve them. A classroom experiment reveals that the revised questions, informed by the behaviors of generative students, become "less difficult" as intended.



\begin{acks}
    This work was funded by NSF Grants DRL-2335975,and IIS-2302564. The findings and conclusions expressed in this material are those of the author(s) and do not necessarily reflect the views of the National Science Foundation.
\end{acks}

\bibliographystyle{ACM-Reference-Format}
\bibliography{reference}

\end{document}